\documentclass{cpbtex}
\graphicspath{{Figs/}}

\usepackage[dvipdfmx]{graphicx}

\begin{document}
\begin{CJK*}{UTF8}{gbsn}


\title{Variational approximation method for the long-range force transmission in biopolymer gels\thanks{Project supported by the National Natural Science Foundation of China (Grant No.~12004082), by Guangdong Province Universities and Colleges Pearl River Scholar Funded Scheme (2019).}}


\author{Haiqin Wang(王海钦)$^{1,2}$  \ and  \ Xinpeng Xu(徐新鹏)$^{1,2}$\thanks{Corresponding author. E-mail:xu.xinpeng@gtiit.edu.cn}\\
$^{1}${\it Physics Program, Guangdong Technion - Israel Institute of Technology,}
\\
{\it 241 Daxue Road, Shantou, Guangdong, China, 515063} 
\\
$^{2}${\it Technion-Israel Institute of Technology, Haifa, Israel, 32000}}  


\date{\today}
\maketitle

\begin{abstract}
The variational principle of minimum free energy (MFEVP) has been widely used in the study of soft matter statics. MFEVP can be used not only to derive equilibrium equations (including both bulk equations and boundary conditions), but also to develop direct variational methods (such as Ritz method) to find approximate solutions to these equilibrium equations. In this work, we applied these variational methods to study long-range force transmission in nonlinear elastic biopolymer gels. We showed that the slow decay of cell-induced displacements measured experimentally for fibroblast spheroids in three-dimensional fibrin gels can be well explained by variational approximations based on the three-chain model of biopolymer gels. 
\end{abstract}



\section{Introduction}\label{Sec:Intro} 

Cells in animal tissues are surrounded by extracellular matrix (ECM)\ucite{Alberts2007}. The physical, topological, and biochemical composition of ECM are not only complex, tissue-specific, but are also markedly heterogeneous\ucite{Alberts2007}. However, in most \emph{in vitro} experiments, only one or two types of constituting proteins (such as collagen, fibrin, and elastin) are extracted from ECM to form cross-linked biopolymer gels\ucite{Alberts2007,Janmey2013,Koenderink2019}. A striking feature of these reconstituted biopolymer gels is their asymmetric elastic response to extension (or shear) and compression\ucite{MacKintosh2014,GardelMacKintosh2004,Lubensky2005}. They stiffen under small shear stresses (or small strains about  5\%-10\%) with their elastic moduli increasing as the 3/2-power-law of the applied stress. In contrast, biopolymer gels soften when compressed upon a small amount where they almost lose resistance to shear stress completely\ucite{Janmey2016,MacKintoshJanmey2016}. Such nonlinear elasticity of fibrous biopolymer gels can be attributed to the microstructural nonlinearities of the constituent semiflexible filaments which stiffen (due to inextensibility) under extension and soften (due to buckling) under compression\ucite{MacKintosh2014,Xu2017PRE,Meng2016,Meng2017,Xu2020}. 

When cells are embedded in such fibrous biopolymer gels, they pull on the gel and produce non-equilibrium forces by myosin motors that consume adenosine triphosphate (ATP). Recent experiments have shown that the displacements and structure changes induced by cells adhered to nonlinear biopolymer gels can reach a distance of tens of cell diameters away\ucite{Harris1981,Janmey2009,Shenoy2016,NotbohmLesman2015,Korff1999}, in contrast to the distance of several cell diameters reached by displacements induced by cells on linear synthetic gels. Such phenomena support long-range cell-cell mechanical communication, a process that can mechanically couple distant cells and coordinate processes such as capillary sprouting\ucite{Korff1999} and synchronous beating\ucite{Tzlil2016}. The long-range transmission of cellular forces is usually attributed to the unique nonlinear mechanics and fibrous nature of the biopolymer gels\ucite{Janmey2013,Janmey2009,MacKintosh2014,Xu2015PRE,Shenoy2014a,Shenoy2014b,Lenz2019,Xu2020,Xu2020BJ}. In this work, we show that a direct variational analysis based on the three-chain continuum model of biopolymer gels\ucite{Xu2015PRE,Xu2020} and the variational principle of minimum free energy (MFEVP) can explain the slow power-law decay of cell-induced displacements that is measured experimentally for fibroblast spheroids embedded in three-dimensional fibrin gels\ucite{NotbohmLesman2015}.  


This paper is organized as follows. After a brief introduction in Sec.~1, we introduce general variational methods in Sec.~2 for static problems in elastic materials. In Sec.~3, we then apply the variational methods to study the transmission of internal forces applied by a spherically contracting cell in a three-dimensional biopolymer gel. In Sec.~4, we summarize our major results and give some remarks. 

\section{Variational methods for elastic materials}\label{Sec:VarMeth}

Variational principles have been widely used in the continuum modeling of soft and biological matter such as the variational principle of minimum free energy (MFEVP) for static problems\ucite{Sam2018} and Onsager's variational principle (OVP) for dynamic problems\ucite{Xu2021,Doi2016,Doi2020,ChunLiu2020,QWang2020}. In this work, we apply MFEVP to study the transmission of internal cellular forces in elastic biopolymer gels. 

\subsection{Variational formulation for the continuum modeling of elastic biopolymer gels}\label{Sec:ActSolid-Energy}

We consider a general elastic material, in which the total free energy functional can be written as 
\begin{equation}\label{Eq:VarMeth-Ftot}
{\cal F}_{\rm t}[\bm{u}({\bm{r}})] = {\cal F}_{\rm e}[\bm{u}({\bm{r}})]
-\int  d\bm{r} f_{i} u_{i}-\oint dA \sigma_{{\rm s} i} u_{i}.
\end{equation}
Here $\bm{u}({\bm{r}})$ is the displacement field, ${\cal F}_{\rm e} [\bm{u}({\bm{r}})]= \int d\bm{r} F_{\rm e}(\bm{u}({\bm{r}}))$ is the deformation energy functional of the elastic material with $F_{\rm e}(\bm{u}({\bm{r}}))$ being the material-dependent energy density. 
$\bm{f}$ and $\bm{\sigma}_{\rm s}$ are the force densities applied in the bulk and at the surfaces, respectively. 
Minimization of ${\cal F}_{\rm t}$ with respect to $\bm{u}({\bm{r}})$ gives the bulk equilibrium (or force balance) equations and boundary conditions:
\begin{subequations}\label{Eq:VarMeth-EquilEqn12}
\begin{align}
&\partial_j\sigma_{ij}^{\rm e}+f_i=0, \\
-\hat{n}_j \sigma_{ij}^{\rm e}+\sigma_{{\rm s} i}&=0, 
\quad {\rm or} \quad u_i=u_{{\rm s}i},
\end{align}
\end{subequations}
respectively. Here $\sigma^{\rm e}_{ij}$ is the elastic stress tensor, generally satisfying $\delta {\cal F}_{\rm e}=\int d\bm{r}\sigma^{\rm e}_{ij} \delta \epsilon_{ij}$ for both linear and nonlinear materials with $\epsilon_{ij}=\frac{1}{2}(\partial_i u_j +\partial_j u_i)$ being the strain tensor\ucite{Landau1986}. $u_{{\rm s}i}$ is some given displacement at the material boundary. 

The explicit form of the deformation energy density ${\cal F}_{\rm e}$ depends on the structure, interactions, and properties of the constituents of the elastic material. For biopolymer gels that are composed of crosslinked stiff biopolymer filaments, several energy forms have been proposed from some continuum models analogous to rubber elasticity\ucite{Treloar1975,MacKintosh2014,PalmerBoyce2008,Xu2015PRE,Meng2017,Xu2020}, \emph{e.g.},  1-chain sphere models, 3- and 8-chain cubic lattice models, and 4-chain tetrahedra model. In these models, the macroscopic elastic deformation energy of the gel is obtained by adding the free energy of individual blocks (or material elements) that are deformed affinely. Here we use the simple 3-chain model that can describe the stretch-stiffening and compressive-softening nonlinearities of biopolymer gels at \emph{small affine} deformation\ucite{PalmerBoyce2008,Xu2015PRE,Meng2017,Xu2020}. In this case, the deformation energy density $F_{\rm e}$ takes the following form\ucite{PalmerBoyce2008,Xu2015PRE,Xu2020} 
\begin{equation}\label{Eq:ActSolid-NonlinFe}
F_{\rm e}=\mu_0\left[ \epsilon_1^2 \left(\frac{6}{5}\frac{1}{1- \epsilon_1/\epsilon_{\rm s}} -\frac{1}{5} \right)+ \tilde{\rho}_2\epsilon_2^2+ \tilde{\rho}_3\epsilon_3^2\right] + \frac{\tilde{K}}{2}(\epsilon_1+\epsilon_2+\epsilon_3)^2
\end{equation}
in terms of the three principal strain components $\epsilon_i$ ($i=1,2,3$) and for the special case of $\epsilon_{1}>0$ and $\epsilon_{2},\epsilon_{3}\leq 0$. 
Here $\tilde{K}\equiv K_0-2\mu_0/3$ is the modified bulk modulus\ucite{Xu2015PRE,Xu2020} with $\mu_0$ and $K_0$ being the linear shear and bulk moduli, respectively, which are related to Young's modulus and Poisson's ratio by $\mu_0={E_0}/{2(1+\nu_0)}$, and $K_0={E_0}/{3(1-2\nu_0)}$. The dimensionless parameter $\tilde{\rho}_{2,3}$ is given by $\tilde{\rho}_{2,3}=1-(1-\rho(\epsilon_{2,3})) (1+\epsilon_b/\epsilon_{2,3})^2$ with $\rho(\epsilon)=\rho_0+(1-\rho_0)\Theta(\epsilon +\epsilon_b)$, $\Theta$ being the Heaviside step function, and $0 \leq \rho_0 \ll 1$.  
Note that the two strain parameters $\epsilon_{\rm b}>0$ and $\epsilon_{\rm s}>0$ denote the magnitude of the threshold strains for the presence of nonlinear compressive softening and stretch stiffening, respectively.
Furthermore, from $\sigma_{i}^{\rm e}=\partial F_{\rm e}/\partial \epsilon_{i}$ we obtain the three principal stress components: 
\begin{subequations}\label{Eq:ActSolid-NonlinSigmaE}
\begin{align}
\sigma_1^{\rm e}=\frac{6}{5}\mu_0\epsilon_s \left[(1-\epsilon_1/\epsilon_s)^{-2}-1 -{\epsilon_1}/{3\epsilon_s} \right] + \tilde{K}(\epsilon_1+\epsilon_2+\epsilon_3), \\
\sigma_{2,3}^{\rm e}=\frac{}{}2\mu_0\left[\epsilon_{2,3}-(1-\rho)(\epsilon_{2,3}+\epsilon_b)\right]+ \tilde{K}(\epsilon_1+\epsilon_2+\epsilon_3).
\end{align}
\end{subequations}

\subsection{Variational methods of approximation: Ritz method}\label{Sec:VarMeth-Ritz}

In the above, we have shown that variational principles such as MFEVP provide an equivalent (and more convenient) applications of vectorial governing (force balance or Euler-Lagrange) equations. However, variational principles should not be regarded only as a mathematical substitute or reformulation of force balance equations. They also provide some powerful variational methods of finding approximate solutions to these equations, \emph{e.g.}, Ritz method and the least-squares method\ucite{Reddy2017}. In these variational methods, some trial solutions to the problem are assumed where the state variable functions are taken as combinations of some simple functions with a small number of adjustable parameters. 
Then the total free energy that is a functional of state variables can be integrated over space and reduces to a function of these adjustable parameters. Correspondingly, the functional minimization of the free energy with respect to state variables reduces to a function minimization with respect to a small number of parameters, which gives approximate solutions of the static problem and determines the equilibrium properties of the system. Such methods simplify the static boundary value problem significantly; they bypass the derivation of the governing equilibrium equations and goes directly from the variational statement to an approximate solution of the static problem. These simplified solution methods are, therefore, called direct variational methods or variational methods of approximation\ucite{Reddy2017}.  

Note that the trial solutions can be either \emph{completely empirical} arising from experiences gained in systematic numerical analysis or experimental measurements\ucite{Doi2015,Doi2020}, or assumed to be \emph{linear combinations of a finite set of basis functions} such as algebraic and trigonometric polynomials\ucite{Reddy2017}. The latter choice of trial solutions is usually known as Ritz method, in which the trial solution can be approximated to arbitrary accuracy by a suitable linear combination of a sufficiently large set of basis functions. However, we would like to emphasize that no matter what forms the trial solutions are assumed to be; they have to satisfy the specified essential boundary conditions (not need to satisfy the natural boundary conditions explicitly, because they are included intrinsically in the variational statement).

To be specific in elastic materials, the only state variable is the displacement field $\bm{u}(\bm{r})$ and the total free energy functional is ${\cal F}_{\rm t}={\cal F}_{\rm t}[\bm{u}(\bm{r})]$. The trial approximate solution of $\bm{u}(\bm{r})$ can be \emph{completely empirical}, for example, taking the form of $\bm{U}(\bm{r};\bm{c})$, which satisfies the specified essential boundary conditions and is parameterized by $N$ as yet unknown (independent) parameters $\bm{c}=(c_{1}, c_{2}, \ldots, c_{N})$. Besides, in the Ritz method, we seek a more explicit approximation $\bm{U}(\bm{r};\bm{c})$, for a fixed and pre-selected $N$, in the finite series form of $\bm{U}(\bm{r};\bm{c}) =\sum_{i=1}^{N} c_{i} \bm{\phi}_{i}(\bm{r})+\bm{\phi}_{0}(\bm{r})$, 
in which $\bm{\phi}_{i}(\bm{r})\, (i=1,2, \ldots, N)$ are basis functions and $c_{i}$ are the unknown (independent) parameters. Here $\phi_{0}(x)$ is chosen to satisfy the specified essential boundary conditions of the problem, and $\phi_{i}$ must be continuous, linearly independent, and satisfy the homogeneous form of the specified essential boundary conditions.  
Substituting either form of the approximate trial solutions $\bm{U}(\bm{r};\bm{c})$ into the total free energy functional ${\cal F}_{\rm t}[\bm{u}(\bm{r})]$, we obtain (after carrying out the integration with respect to $\bm{r}$): ${\cal F}_{\rm t}={\cal F}_{\rm t}(\bm{c})$. The parameters $\bm{c}$ are then determined by minimizing ${\cal F}_{\rm t}$ with respect to $\bm{c}$:
\begin{equation}\label{Eq:VarMeth-FtMin}
\frac{\partial {\cal F}_{\rm t}}{\partial c_{i}}=0 \quad \text { for } i=1,2, \ldots, N
\end{equation}
which represents a set of $N$ linear equations among $c_{1}, c_{2}, \ldots, c_{N}$, whose solution together with the above trial solution yields the approximate solution $\bm{U}(\bm{r})$. This completes the description of the variational methods of approximation.  

\section{Transmission of forces induced by a spherically contracting cell in biopolymer gels}\label{Sec:ForceTrans} 

Now we use the above variational methods to study the decay of displacements $u(r)$ induced by a spherically contracting cell that is well adhered to a 3D nonlinear elastic biopolymer gel\ucite{Xu2015PRE,Xu2020,Yair2021}. Particularly, we show that an approximate one-parameter trial solution of power-law form, $u\sim r^{-n}$, can explain the peculiarly slow decay of cell-induced displacements with $n\approx 0.52$, which is measured in previous experiments of fibroblast cells embedded in 3D fibrin gels\ucite{NotbohmLesman2015}.
 
\begin{center}
\includegraphics[width=0.85\textwidth]{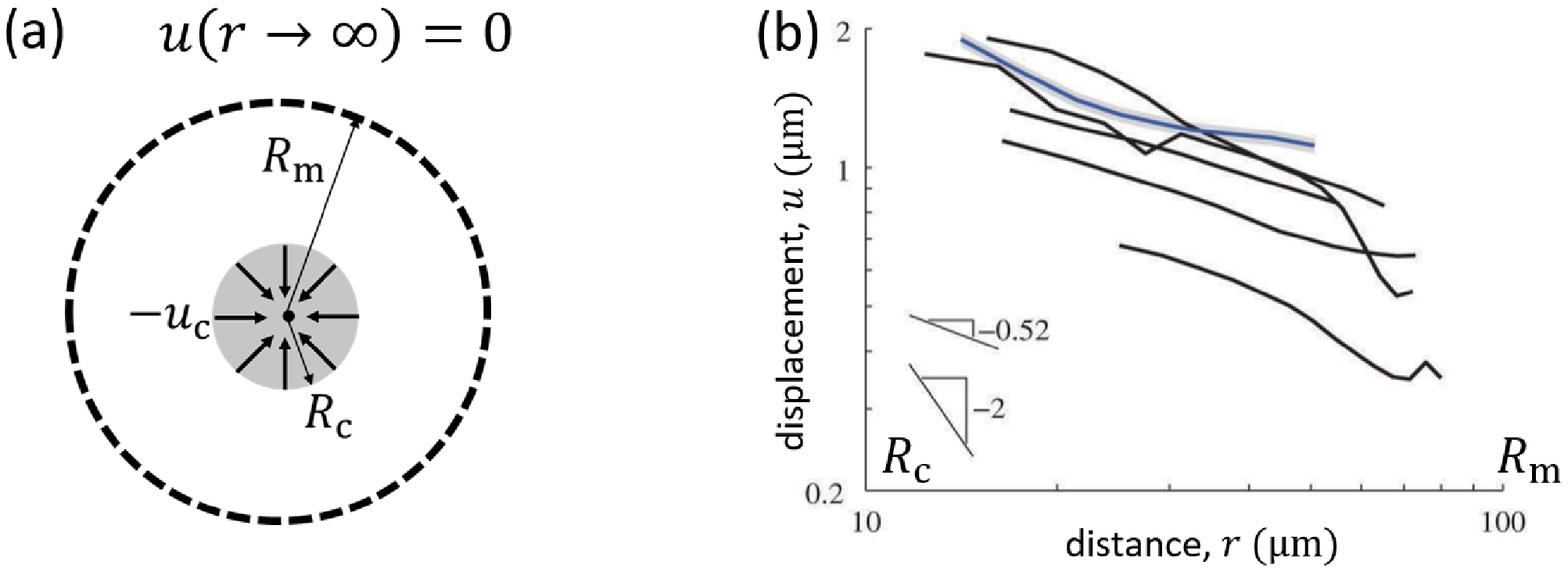}\\[5pt]
\parbox[c]{15.0cm}{\footnotesize{\bf Fig.~1.} (color online) 
  (a) Schematic illustration of a spherically contracting cell of radius $R_{\rm c}$ in an infinite elastic matrix. The cell applies a displacement $-u_{\rm c}$ at its boundary and $u_{\rm c}>0$ for contractile cells. (b) The decay of displacement $u(r)$ induced by a spherically contracting cell is measured in experiments where spherical fibroblast cells are embedded in 3D fibrin gels. An effective power-law $u\sim r^{-n}$ with exponent $n\approx 0.52$ is identified. Reproduced from Notbohm \emph{et al.}\ucite{NotbohmLesman2015} with permission from Royal Society. We consider the decay of displacement in the range of $R_{\rm c}\leq r<R_{\rm m}$. }
  \label{Fig:Exp}
\end{center}

An adherent cell can apply active contractions to their surrounding matrix. In this sense, the cell can be modeled as a contractile force dipole\ucite{Sam2013a}.
In the simplest case, we here consider the decay of displacements that are induced by a spherically contracting cell in a 3D infinite extracellular matrix\ucite{Sam2012b,Sam2015,Xu2015PRE} as done in experiments and schematically shown in Fig.~1(a). In this case, the active cell contraction can be characterized by a boundary condition of fixed radial displacement $-u_c$ at the cell boundary $r=R_c$ as shown in Fig.~1(a), \emph{i.e.},
\begin{equation}\label{Eq:App2-CellBcs}
\bm{u}(r=R_c)=-u_c{\bm{\hat{r}}}
\end{equation}
with $u_c>0$ for contractile cells.  
In spherical coordinates ($r,\theta,\varphi$), the displacement vector is then given by $\bm{u}= u(r) \hat{\bm {r}}$, the non-zero components of the strain tensor $\epsilon_{ij}$ are
\begin{equation}\label{Eq:App2-Strain} \epsilon_{rr}=u^{\prime} \equiv \frac{d u}{d r}>0, \quad \epsilon_{\theta \theta}=\epsilon_{\varphi \varphi}=\frac{u}{r}<0. 
\end{equation} 
and the total energy is then given from Eq.~(\ref{Eq:VarMeth-Ftot}) by
\begin{equation}\label{Eq:App2-Ft}
{\cal F}_{\rm t}[u(r)] = \int_0^R dr 4 \pi r^2 F_{\rm e}(u(r)) -4\pi R_{\rm c}^2 \sigma_{{\rm c}} u(r=R_{\rm c}).
\end{equation} 
Here the deformation energy density $F_{\rm e}$ for biopolymer gels is given in Eq.~(\ref{Eq:ActSolid-NonlinFe}), and $\sigma_{\rm c}<0$ is the contractile stress applied by the living cell at its boundary. From the minimization of ${\cal F}_{\rm t}$ with respect to $u(r)$, we obtain the bulk equilibrium condition\ucite{Sam2012b,Sam2015,Xu2015PRE,Shenoy2014b,Lenz2019}
\begin{equation}\label{Eq:App2-EquilEqn}
\frac{d \sigma^{\rm e}_{\rm rr}}{d r}+\frac{2}{r}\left(\sigma^{\rm e}_{\rm rr}-\sigma^{\rm e}_{\rm \theta \theta}\right)=0,
\end{equation}
and the boundary condition in Eq.(\ref{Eq:App2-CellBcs}): $u(r=R_{\rm c}) =- u_{\rm c}$, which is supplemented with the natural boundary condition: $u(r\rightarrow \infty) = 0$. Here $\sigma^{\rm e}_{ij}=\partial F_{\rm e}/\partial \epsilon_{ij}$ is the elastic stress and its principal components are given in Eq.~(\ref{Eq:ActSolid-NonlinSigmaE}). Note that the cellular stress $\sigma_{\rm c}$ can be calculated by $\sigma_{{\rm c}}=\frac{1}{4\pi R_{\rm c}^2}\frac{\partial {\cal F}_{\rm t}}{\partial u_{\rm c}}$. 

\subsection{Force transmission in biopolymer gels under linear limits}\label{Sec:App2-Linear}

We first calculate the decay of displacements or force transmission in the following two linear limits. 

\emph{Force transmission in linear isotropic limit: $\epsilon_1/\epsilon_s\ll 1$ and $|\epsilon_{2,3}|/\epsilon_b \ll 1$}. In this limit, the elastic energy density in Eq.~(\ref{Eq:ActSolid-NonlinFe}) reduces to
\begin{equation}\label{Eq:App2-LinIsoFe}
F_{\rm e}=\frac{2}{3} \mu_0\left(u'-\frac{u}{r}\right)^{2}+\frac{K_0}{2}\left(u'+ \frac{2u}{r}\right)^{2}.
\end{equation}
which takes the same form of $F_{\rm e}$ for linear isotropic materials\ucite{Landau1986,Xu2020} if substituting $\mu_0={E_0}/{2(1+\nu_0)}$ and $K_0={E_0}/{3(1-2\nu_0)}$.
Using Ritz-type method explained in Sec.~2.2, we take the empirical trial solution of the power-law form, $u=-u_c(R_c/r)^n$, which satisfies the essential boundary conditions at $r=R_{\rm c}$ and $r\rightarrow \infty$, and has only one undetermined parameter $n$. In this case, the energy density $F_{\rm e}$ can be integrated in space $R_{\rm c}\leq r<\infty$ and then the deformation energy functional $\mathcal{F}[u(r)]=\int_{R_{\rm c}}^{\infty} dr 4 \pi r^2 F_{\rm e}(u(r))$ becomes a function of $n$ as
\begin{equation}\label{Eq:App2-LinIsoFn}
\mathcal{F}(n)=\int_{R_{\rm c}}^{\infty} F_{\rm{e}}(u(r)) 4 \pi r^{2} d r=\frac{4 \pi R_{\rm c} u_{\rm c}^{2}}{2 n-1}\left[\frac{2}{3} \mu_0(n+1)^{2}+\frac{K_0}{2}(n-2)^{2}\right].
\end{equation}
Minimization of $\mathcal{F}(n)$ with respect to $n$ gives $n=2$. That is, the solution is $u=-u_c(R_c/r)^2$, which is the exact solution of the boundary value problem in linear isotropic materials. 

\emph{Force transmission in linear transtropic limit: $\epsilon_1/\epsilon_s \ll 1$ and $|\epsilon_{2,3}|/\epsilon_b >1$}. In this case, the elastic energy density in Eq.(\ref{Eq:ActSolid-NonlinFe}) reduces to
\begin{equation}\label{Eq:App2-LinAnisoFe}
F_{\rm e}=\mu_0\left( \epsilon_1^2 + \rho_0\epsilon_2^2+ \rho_0\epsilon_3^2\right) + \frac{\tilde{K}}{2}(\epsilon_1+\epsilon_2+\epsilon_3)^2,
\end{equation} 
which can be casted into the general energy form in linear transtropic materials\ucite{Lekhnitskii1981,Xu2020} as 
\begin{equation}\label{Eq:App2-LinAnisoFe2}
F_{\rm e}=\frac{1}{2}C_{1}u'^{2}+8 C_{2}\left(\frac{u}{r}\right)^{2}+4 C_{4}\frac{u'u}{r}
\end{equation} 
with coefficients given by
\begin{equation}\label{Eq:App2-LinAnisoC}
C_1=2\mu_0+\tilde{K}, \quad
C_2=\frac{1}{4}(\rho_0\mu_0+\tilde{K}),  \quad
C_3=\frac{1}{2}\rho_0\mu_0, \quad
C_4=\frac{1}{2}\tilde{K}, 
\end{equation} 
and the bulk modulus given by $K_0=\tilde{K}+\frac{2}{9}(1+2\rho_0)\mu_0$.
We take the same empirical trial solution as above, $u=-u_c(R_c/r)^n$, and then the deformation energy $\mathcal{F}$ functional also becomes a function of $n$ as
\begin{equation}\label{Eq:App2-LinAnisoFn}
\mathcal{F}(n)=\int_{R_{0}}^{\infty} F_{\rm e}(u(r)) 4 \pi r^{2} d r=\frac{4\pi R_{\rm c}u_{\rm c}^2}{2n-1}\left(\frac{1}{2}C_1n^2-4C_4n+8C_2\right).
\end{equation}
Minimization of $\mathcal{F}(n)$ with respect to $n$ gives $n=\frac{1}{2}(1+\sqrt{1+8 {\cal B}})$, which is also the exact solution to the equilibrium equations in linear transtropic materials. Here ${\cal B}=(8C_2-2C_4)/C_1$ and using Eq.(\ref{Eq:App2-LinAnisoC}), we get ${\cal B}=(2\rho_0\mu_0 +\tilde{K})/(2\mu_0+\tilde{K})$.

\begin{center}
\includegraphics[width=0.9\textwidth]{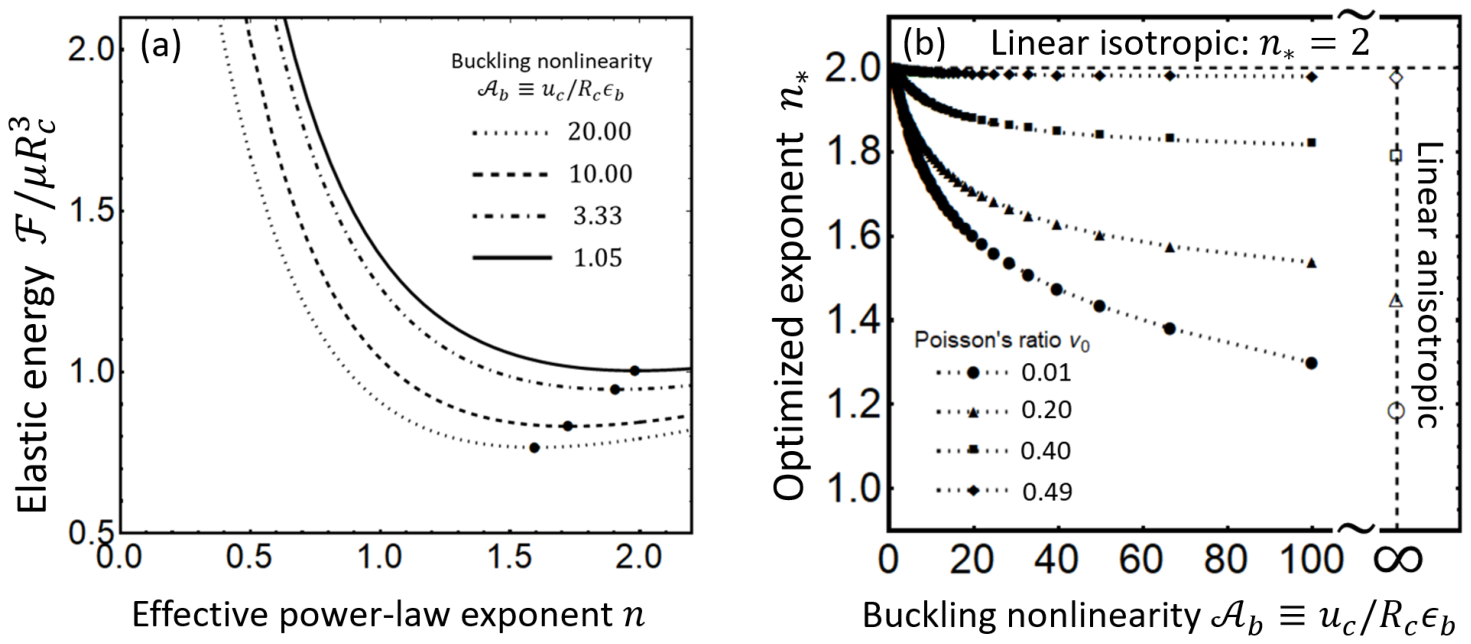}\\[5pt]
\parbox[c]{15.0cm}{\footnotesize{\bf Fig.~2.} (color online) 
An approximating power-law decay of cell-induced displacements in nonlinear gels of compressive-softening but no stretch-stiffening. (a) Elastic deformation energy $\cal F$ as a function of the power-law exponent $n$. Several different nonlinearity parameters ${\cal A}_{\rm b}\equiv u_{\rm c}/R_{\rm c} \epsilon_{\rm b}$ have been chosen. The black dots in the curves denote the optimized exponent $n_*$ that corresponds to minimum energy. Here we take $\nu_0=0.01$, $u_c/R_c=0.2$, $\rho_0=0.1$. (b) The optimized exponent $n_*$ is plotted as a function of ${\cal A}_{\rm b}$ for different Poisson's ratios. Here we also take $u_c/R_c=0.2$ and $\rho_0=0.1$.}
\label{Fig:App2-Nonlin1}
\end{center} 

\subsection{Force transmission in nonlinear biopolymer gels}\label{Sec:App2-Nonlin}

We now calculate the decay of displacements in nonlinear elastic biopolymer gels. 

\emph{Force transmission in nonlinear gels with compressive-softening but no stretch-stiffening: $\epsilon_{\rm s} \rightarrow \infty$.} In this limit, the elastic energy density in Eq.(\ref{Eq:ActSolid-NonlinFe}) reduces to
\begin{equation}\label{Eq:App2-NonlinFe}
F_{\rm{e}}(u(r))=\mu_0\left(\epsilon_{1}^{2}+2 \tilde{\rho} \epsilon_{2}^{2}\right)+\frac{1}{2}\left(K_0-\frac{2}{3} \mu_0\right)\left(\epsilon_{1}+2 \epsilon_{2}\right)^{2}, 
\end{equation}
with $\tilde{\rho}=1-\left[1-\rho\left(\epsilon_{2}\right)\right]\left(1+{\epsilon_{\rm b}}/{\epsilon_{2}}\right)^{2}$ and $\rho\left(\epsilon_{2}\right)=\rho_{0}+\left(1-\rho_{0}\right) \Theta\left(\epsilon_{2}+\epsilon_{\rm b}\right)$. 

To use the variational method of approximation, we take the same empirical trial solution as above: $u=-u_c(R_c/r)^n$. Therefore, the angular strain component $\epsilon_{2}=u/r$ equals to $-\epsilon_{\rm b}$ at $r=R_{\rm b} \equiv R_{\rm c}\left(u_{c} / R_{c} \epsilon_{\rm b}\right)^{\frac{1}{n+1}}$, which divides the matrix into two regions. In the far field ${r} >{R}_{\rm b}$, we have $\epsilon_{2}+\epsilon_{\rm b}>0$ and hence $\rho(\epsilon_{2})=1$, $\tilde{\rho}=1$, that is, the matrix behaves as a linear and isotropic material. In the near field $R_{\rm c} \leq r \leq {R}_{\rm b}$, we have $\epsilon_{2}+\epsilon_{\rm b}<0$ and hence $\rho(\epsilon_{2})=\rho_{0}$, $\tilde{\rho}=1-\left(1-\rho_{0}\right)\left(1+\frac{\epsilon_{\rm b}}{\epsilon_{2}}\right)^{2}$, that is, the matrix is linear transtropic. The deformation energy functional then becomes a function of $n$ as
\begin{equation} \label{Eq:App2-NonlinFn}
\mathcal{F}\left(n; \epsilon_{\rm b}, \nu_{0}, \rho_{0}\right)=
R_{\rm c}^{3} \int_{{R}_{\rm c}}^{{R}_{\rm b}} F_{\rm{e}}[u(r)] 4 \pi {r}^{2} d r
+R_{\rm c}^{3} \int_{{R}_{\rm b}}^{{R}_{\rm m}} F_{\rm{e}}[u(r)] 4 \pi r^{2} d r.
\end{equation} 
Here we do the integral over a finite matrix region, $R_{\rm c} \leq r \leq {R}_{\rm m}$, take ${R}_{\rm m}=10 R_{\rm c}>{R}_{\rm b}$, and we plot the energy $\mathcal{F}(n)$ in Fig.~2(a) for different nonlinearity parameters, ${\cal A}_{\rm b}=u_{\rm c}/R_{\rm c} \epsilon_{\rm b}$. We then minimize $\mathcal{F}(n)$ numerically with respect to $n$ and obtain the optimized power-law exponent $n_*$ as a function of $\epsilon_{\rm b}$, $\nu_{0}$, and $\rho_{0}$. In Fig.~2(b), $n_*$ is plotted as a function of ${\cal A}_{\rm b}$ for different Poisson's ratios. In the limit of ${\cal A}_{\rm b}\to \infty$, the biopolymer gels behave simply as linear anisotropic material and the corresponding $n_*$ are marked by open symbols. As $\nu_0$ increases to the incompressible limit $\nu_0=0.5$, $n_*$ approaches to its value in linear isotropic materials, $n_*=2$. 

\begin{center}
\includegraphics[width=0.9\textwidth]{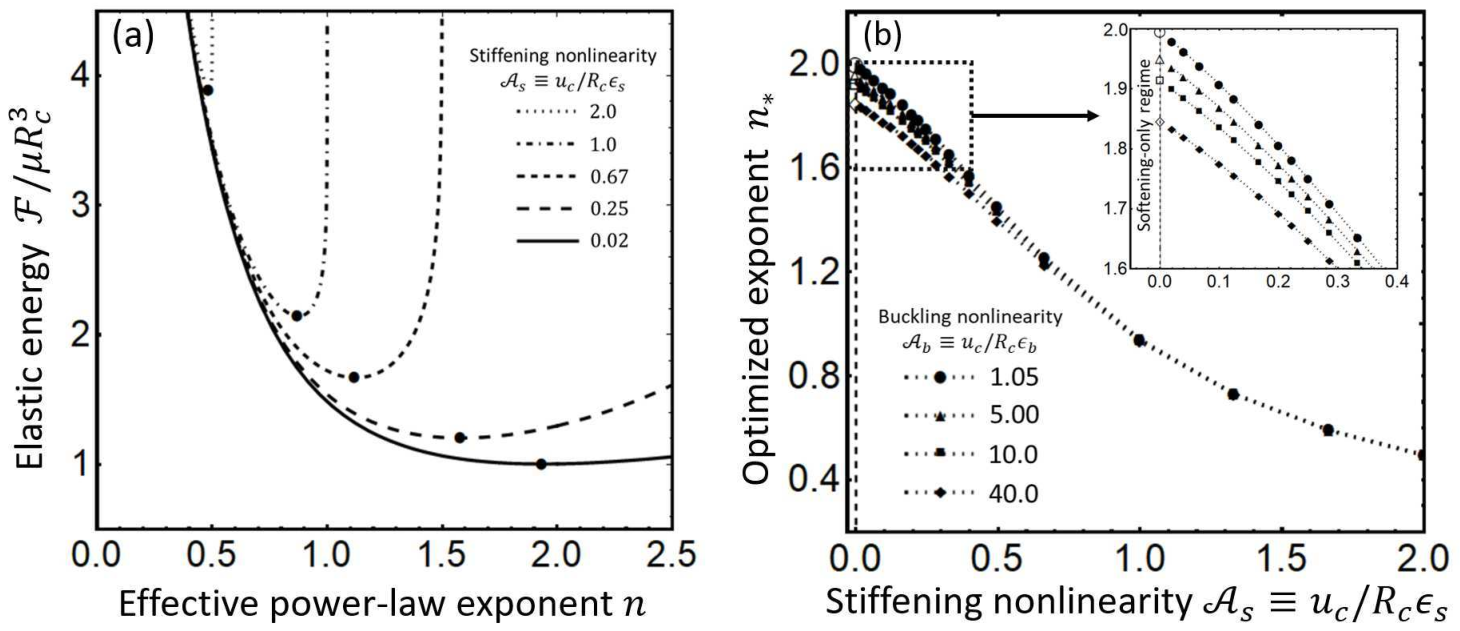}\\[5pt]
\parbox[c]{15.0cm}{\footnotesize{\bf Fig.~3.} (color online) 
An approximating power-law decay of cell-induced displacements in nonlinear gels with both compressive-softening and stretch-stiffening. (a) Elastic deformation energy $\cal F$ as a function of the power-law exponent $n$. Several different stiffening nonlinearity parameters ${\cal A}_{\rm s}\equiv u_{\rm c}/R_{\rm c} \epsilon_{\rm s}$ have been chosen. The black dots in the curves denote the optimized exponent $n_*$ that corresponds to minimum energy. Here we take $\nu_0=0.4$, $u_c/R_c=0.2$, $\rho_0=0.1$, and $\mathcal{A}_{\rm b}=2.0$ (with $\epsilon_b=0.1$). (b) The optimized exponent $n_*$ is plotted as a function of ${\cal A}_{\rm s}$ for different buckling nonlinearity parameter $\mathcal{A}_{\rm b}$. Here we also take $u_c/R_c=0.2$ and $\rho_0=0.1$. }
\label{Fig:App2-Nonlin2} 
\end{center}

\emph{Force transmission in biopolymer gels with both stretch-stiffening and compressive-softening.} 
In general, one has to consider the full elastic energy density $F_{\rm e}$ in the form of Eq.(\ref{Eq:ActSolid-NonlinFe}), which takes into account of both stretch-stiffening and compressive softening. In this case, we still take the approximate empirical trial solution, $u=-u_c(R_c/r)^n$, and numerically integrate the deformation energy functional $\mathcal{F}[u(r)]$ over a finite matrix region, $R_{\rm c} \leq r \leq {R}_{\rm m}$, and obtain an energy function of $n$: $\mathcal{F}\left(n; \epsilon_{\rm s}, \epsilon_{\rm b}, \nu_{0}, \rho_{0}\right)$. 
In Fig.~3(a), we plot the energy $\mathcal{F}(n)$ for different nonlinearity parameters, ${\cal A}_{\rm s}=u_{\rm c}/R_{\rm c} \epsilon_{\rm s}$ where we take ${R}_{\rm m}=10 R_{\rm c}>{R}_{\rm b}$ as measured in experiments\ucite{NotbohmLesman2015}. We then minimize $\mathcal{F}(n)$ numerically with respect to $n$ and obtain the optimized power-law exponent $n_*$ as a function of $\epsilon_{\rm s}$, $\epsilon_{\rm b}$, $\nu_{0}$, and $\rho_{0}$. In Fig.~3(b), we plot $n_*$ as a function of ${\cal A}_{\rm s}$ for various ${\cal A}_{\rm b}$. In the limit of ${\cal A}_{\rm s}\to 0$, the biopolymer gels behave in the same way as nonlinear gels with only compressive-softening (as shown in Fig.~2) and the corresponding $n_*$ are marked by open symbols in Fig.~3(b). 

\begin{center}
\includegraphics[width=0.55\textwidth]{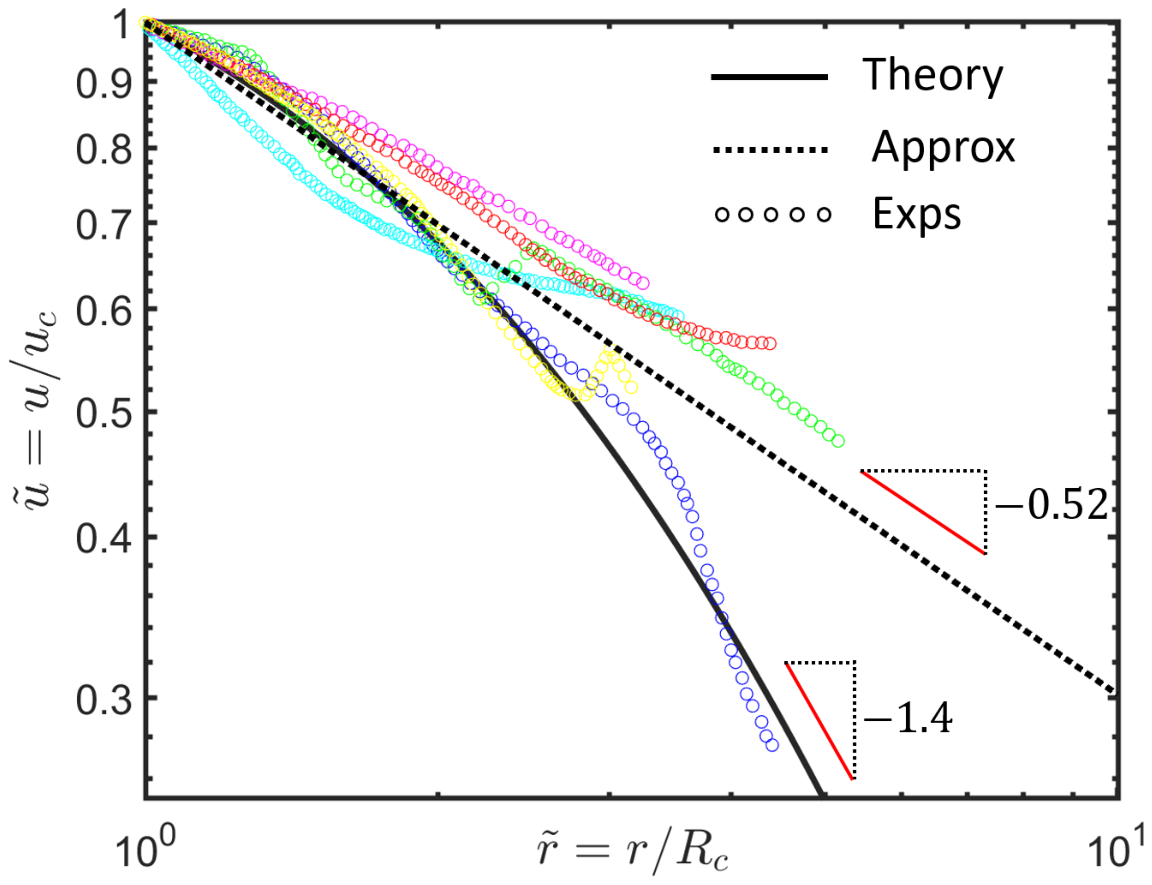}\\[5pt]
\parbox[c]{15.0cm}{\footnotesize{\bf Fig.~4.} (color online) 
The decay of displacement induced by the spherically contracting cell in nonlinear biopolymer gels with both compressive-softening and stretch-stiffening. The approximating power-law $u=-u_c(R_c/r)^{-0.52}$ (shown as dashed lines) measured in experiments\ucite{NotbohmLesman2015} is fitted here by choosing $\nu_0=0.1, \, \rho_0=0.1, \, {\cal A}_{\rm b}=33.3$, and ${\cal A}_{\rm s}=1.76$, which lie in the reasonable range of these parameters measured in experiments. The corresponding numerical solution of the equilibrium equation (\ref{Eq:App2-EquilEqn}) is shown by solid curves. Open symbols with different colors are taken from experiments shown in Fig.~1(b) by Notbohm \emph{et al.}\ucite{NotbohmLesman2015}. }
\label{Fig:App2-Nonlin3}
\end{center}

In addition, we find that the slow power-law decay of displacements induced by contracting cells measured in experiments by Notbohm \emph{et al.}\ucite{NotbohmLesman2015} can be quantitatively explained by the full 3-chain model of biopolymer gels and the above variational approximations. By taking the approximate trial solution, $u=-u_c(R_c/r)^n$, following the above variational calculations, and taking reasonable parameter values measured in experiments, we can obtain $n\approx 0.52$ theoretically, as shown by dotted curves in Fig.~4. Using these parameter values, the equilibrium equations are also solved numerically\ucite{Xu2015PRE} as shown by solid curves in Fig.~4. It can be seen that the theory calculations agree well with experimental data\ucite{NotbohmLesman2015}. 

\section{Conclusions}\label{Sec:Conclude}
Variational methods have been widely used in the modeling and analysis of soft and biological matter. In this work, we used the variational methods that are based on the variational principle of minimum free energy (MFEVP) to study the slow-decay of the displacements induced by a spherically contracting cell in a three-dimensional biopolymer gels. We used the 3-chain model to describe the nonlinear (stretch-stiffening and compressive-softening) elasticity of biopolymer gels. We firstly showed that in the linear limits, the classical scaling laws for the decay of cell-induced displacements in linear isotropic and linear anisotropic elastic medium can both be obtained exactly by the Ritz-type variational methods. We then showed that in general nonlinear biopolymer gels, when relevant physical parameters take reasonable values as that measured in other separate experiments, we can use variational methods to reproduce the scaling law $u\sim r^{-0.52}$ that is measured experimentally for the decay of displacements induced by fibroblast cell spheroid contracting in three-dimensional fibrin gels. This work evidence the validity of continuum modeling in describing fibrous biopolymer gels and deepen our understanding of long-range force transmission in biopolymer gels and extracellular matrix that is essential for efficient matrix-mediated cell-cell communications. 


\section*{Acknowledgment} 
X.X. is supported in part by a project supported by the National Science Foundation for Young Scientists of China (NSFC, No.~12004082), by Guangdong Province Universities and Colleges Pearl River Scholar Funded Scheme (2019), by 2020 Li Ka Shing Foundation Cross-Disciplinary Research Grant (No.~2020LKSFG08A), by Provincial Science Foundation of Guangdong (2019A1515110809), by Guangdong Basic and Applied Basic Research Foundation (2020B1515310005), and by Featured Innovative Projects (No.~2018KTSCX282) and Youth Talent Innovative Platforms (No.~2018KQNCX318) in Universities in Guangdong Province. 

\bibliographystyle{unsrt}  
\bibliography{ForceTrans}  

\end{CJK*}  
\end{document}